\documentclass{appolb}

\usepackage{cite}

\usepackage[final]{graphicx}
\usepackage{bbm}
\usepackage{amsmath,amssymb}

\newcommand{\superN}{\mathcal{N}}

\newcommand{\ve}{{\mathcal E}}

\newcommand{\up}{\,\uparrow\,}

\newcommand{\Integers}{\mathbbm{Z}}
\newcommand{\Reals}{\mathbbm{R}}

\newcommand{\Sphere}{S}
\newcommand{\AdS}{\mathrm{AdS}}

\newcommand{\conj}[1]{\bar{#1}}

\newcommand{\grp}[1]{\mathrm{#1}}

\newcommand{\grSU}{\grp{SU}}

\newcommand{\grSL}{\grp{SL}}

\newcommand{\lrbrk}[1]{\left(#1\right)}

\newcommand{\ket}[1]{\mathopen{|}#1\mathclose{\rangle}}

\begin{document}

\pagestyle{plain}

\title{Finite size giant magnons and interactions%
\thanks{Seminar talk presented at the XLVIII Cracow School of Theoretical Physics, ``Aspects of Duality'', Zakopane, Poland, June 13--22, 2008.}}
\author{Olof~Ohlsson~Sax
\address{Department of Physics and Astronomy, Uppsala University,
Box 803, 751 08 Uppsala, Sweden}}
\maketitle

\begin{abstract}
  Magnon interactions give important contributions to the wrapping
  interactions of the $\superN=4$ spin-chain. Similar effects are
  expected for the finite size corrections to the giant magnon energy
  in $\AdS_5 \times \Sphere^5$. In this paper I review the finite gap
  description of giant magnons and the leading order calculation of
  the finite size corrections to the giant magnon dispersion relation
  for multi-magnon states.

  \vspace{\baselineskip}\noindent
\end{abstract}

\section{Gauge invariant operators and spin-chains}

A major test of the large $N$ AdS/CFT
correspondence~\cite{Maldacena:1997re} is the matching of the spectrum
of the scaling dimensions of single trace operators in $\superN=4$
super Yang-Mills (SYM) and the energies of single strings propagating
in $\AdS_5 \times \Sphere^5$.

Many features of the spectrum of single trace operators can be
understood by treating it as a lattice model. This is most easily
understood in the $\grSU(2)$ sector, which consists of operators built
out of the scalars $Z$ and $Y$. Here the operators can then be mapped
to simple spin-chains. As realized by Minahan and
Zarembo~\cite{Minahan:2002ve}, the one-loop dilation operator in this
sector equals the hamiltonian of the Heisenberg $\grSU(2)$
spin-chain. If we let $Z$ represent spin up and $Y$ represent spin
down, a ferromagnetic ground state of a chain of length $J$ is given
by
\begin{equation}
  \ket{\Omega} = \Tr Z^J = \ket{\up \up \dotsb \up \up}.
\end{equation}
This is a BPS state with classical dimension $E = J$ and vanishing
anomalous dimension.

Non-BPS operators can be built out of this ground state by changing some of
the $Z$:s into $Y$:s. In the spin-chain picture each such impurity
corresponds to a magnon, \ie a single fundamental excitation traveling
along the chain with a certain momentum.

For an infinitely long chain, \ie for $J \to \infty$, the spectrum
consists of states with any number of magnons having arbitrary
momenta. The contribution of each magnon to the scaling dimension, \ie
the dispersion relation of the magnons, is\footnote{The coupling constant $g$
  is related to the 't Hooft coupling $\lambda$ by $g^2 =
  \frac{\lambda}{16\pi^2}$.}~\cite{Beisert:2005tm}
\begin{equation}
\label{eq:disp-rel-gauge}
  \ve = E - J = \sqrt{1 + 16g^2 \sin^2 \frac{p}{2}} \approx 1 + 8g^2 \sin^2 \frac{p}{2}.
\end{equation}

For finite $J$, we need to take into account the periodic
boundary conditions. In an integrable system such as the Heisenberg
spin-chain, this leads to quantization of the momenta, which now
satisfy Bethe equations of the form
\begin{equation}
  e^{i p_i J} = \prod_{j \not= i} S^{-1}(p_j, p_i),
\end{equation}
where $S(p_j, p_i)$ is the two-body S-matrix. Hence the finite $J$
spectrum is highly dependent on the magnon interactions.

To get a gauge independent operator from a spin-chain state, we need
to take a trace. Taking into account the cyclicity of the trace, we
will consider only spin-chain states that are symmetric under shifts of
the spin sites. This leads to the level-matching condition
\begin{equation}
  \label{eq:level-matching}
  \sum p_i = 2\pi m\ ,\qquad m \in \Integers.
\end{equation}

The success of this Bethe ansatz for the $\grSU(2)$ sector relies on
the similarity between the one-loop dilation operator in this sector
and the integrable Heisenberg hamiltonian. Miraculously there are many
indications that even the all-loop dilatation operator of the full
large $N$ $\superN=4$ SYM is integrable. Hence it is possible to
describe the full spectrum of single trace gauge invariant operators
by a Bethe ansatz~\cite{Beisert:2005fw}.

However the Bethe ansatz is only valid for asymptotically long
operators. The range of the interaction terms in the dilatation
operator grow with the loop order. For short operators finite size
corrections are expected to appear in the form of wrapping
interactions. Ambjorn \textit{et~al.\@}~\cite{Ambjorn:2005wa} analyzed
wrapping effects using the Thermodynamic Bethe ansatz (TBA). They
found that for an operator of length $L$, wrapping effects will
generically appear at $L$ loops.

Recently much work has been put into calculating these corrections.
The simplest operator with a non-zero anomalous dimension is the
Konishi operator. The dimension of this operator has been calculated
to four loop order directly from the gauge theory
\cite{Konishi} as well as
using TBA~\cite{Bajnok:2008bm}, and the results of these calculations
were found to agree perfectly.

\section{The spinning point-like string and the giant magnon}

By considering string states in $\Reals \times \Sphere^2 \subset
\AdS_5 \times \Sphere^5$ we find the duals of operators in the
$\grSU(2)$ sector of the gauge theory. The R-charge $J$ now measures
the angular momentum around the sphere. The simplest solution
describes a point-like string spinning around a great circle. This
supersymmetric state with $E = J$ is the dual of the spin-chain ground
state.

Having identified the dual of the ground state the next step is
to search for excitations. Berenstein
\textit{et~al.\@}~\cite{Berenstein:2002jq} considered the limit where
$J, g \to \infty$ with $g/J$ fixed, and showed that the semi-classical
fluctuations around the ground state reproduce the leading order gauge
theory spectrum in this limit.

Another possibility would be to let $J \to \infty$ while keeping $g$
large but fixed. Hofman and Maldacena~\cite{Hofman:2006xt} looked for
classical solutions such that the difference $E - J$ remains
finite. They found a solution given by a world-sheet soliton. In
space-time, the solution describes a string with end-points fixed on
the equator with a constant angular separation $\Delta\varphi$, which
is interpreted as the momentum $p$ of the excitation. Using a
conformal gauge where the $J$ density is constant on the world-sheet,
the world-sheet is infinitely large.  The energy of the excited state
is
\begin{equation}
  \ve = E - J = 4g\sin\frac{p}{2},
\end{equation}
which agrees with the large $g$ limit of the gauge theory result
\eqref{eq:disp-rel-gauge}. The situation is thus very similar to that of small
fluctuations around an infinitely long spin chain.

The giant magnon solution seems to describes an open string in a
closed string theory. However, a single magnon does not describe a
physical configuration. Like in the gauge theory, physical states
satisfy the level-matching condition \eqref{eq:level-matching}. With
an infinite world-sheet we can however relax this condition and
consider a single magnon.\footnote{For finite $J$, single magnon
  states of momentum the form $2\pi \frac{m}{M}$ are physical if
  treated on a $\Reals \times \Sphere^2/\Integers_M$ orbifold,
  see~\cite{Astolfi:2007uz}.}

\subsection{The giant magnon as a finite gap solution}

One approach to deriving the giant magnon dispersion relation is by
constructing a finite gap
solution~\cite{Kazakov:2004qf,Minahan:2006bd,Vicedo:2007rp}. A
classical solution to the equations of motion of the string is
described by a meromorphic function $P(x)$, called the
\textit{quasi-momentum}, defined on a two-sheeted Riemann surface.
The charges of the string depend on the analytical structure of
$P(x)$.  The possible singularities are square root branch cuts
$\mathcal C_k$ and logarithmic branch cuts $\mathcal B_k$,
conventionally referred to as \textit{condensates}.  If $P(x)$ has a
square root branch cut along $\mathcal C_k$ it satisfies
\begin{equation}
  P(x + i\epsilon) + P(x - i\epsilon) = 2\pi n_k, \qquad x \in \mathcal C_k,
  \label{eq:P-discont-eq}
\end{equation}
where $P(x)$ is to be evaluated once on each side of the cut.

The quantum numbers of a particular string configuration can be read
off from the asymptotic behavior of the corresponding quasi-momentum%
\footnote{The finite gap solutions considered here really correspond
  to strings on $\Reals \times \Sphere^3$. The conserved charges $J$
  and $Q$ correspond to the isometries of the three-sphere. For
  solutions made up of giant magnons, $Q$ counts the number of magnon
  constituents, with $Q=1$ corresponding to a fundamental magnon and
  larger $Q$ corresponding to magnon bound states.}%
,
\begin{align}
  P(x) &= \frac{E}{4g} \frac{1}{x \pm 1} + \dotsb, & (x \to \mp 1), \\
  P(x) &= \frac{J - Q}{2 g x} + \dotsb, & (x \to \infty), \\
  P(x) &= p - \frac{J + Q}{2 g} x + \dotsb, & (x \to 0).
\end{align}
Let us now consider a configuration consisting of a single condensate
and make the ansatz
\begin{align}
  P(x) &= \frac{E}{2 g} \frac{x}{x^2-1} + G(x), &
  G(x) &= - i \log \frac{x - X^+}{x - X^-}.
\end{align}

$P(x)$ has a single logarithmic cut between $X^+$ and $X^-$.  The
poles at $\pm 1$ already have the correct residues. To get the right
asymptotic behavior as $x \to 0, \infty$ we have to solve the
equations
\begin{align}
  E - J + Q &= -2ig (X^+ - X^-), \\
  E - J - Q &= 2ig \lrbrk{\frac{1}{X^+} - \frac{1}{X^-}}, \\
  p &= -i \log \frac{X^+}{X^-}.
\end{align}
Solving these equations for $X^{\pm}$ gives
\begin{equation}
  X^\pm = \frac{Q + \sqrt{Q^2 + 16g^2\sin^2 \frac{p}{2}}}{4g} e^{\pm i \frac{p}{2}} \csc \frac{p}{2}.
\end{equation}
Plugging this back, solving for $E - J$ and setting $Q = 1$ gives the dispersion
relation
\begin{equation}
  \ve = \sqrt{1 + 16g^2\sin^2 \frac{p}{2}} \approx 4g\sin\frac{p}{2}.
\end{equation}

To better understand how we arrived at this result, we consider an
alternative finite gap derivation due to Vicedo~\cite{Vicedo:2007rp}.
In this formalism a solution to the classical equations of motion is
encoded in terms of a meromorphic differential on a Riemann
surface. Consider an elliptic, or two gap, solution, \ie a solution
with a genus one Riemann surface. This surface can be described as a
two sheeted surface with two square root branch cuts connecting the
points $X^+$ and $X^-$, and $Y^+$ and $Y^-$.

\begin{figure}
  \centering
  \includegraphics{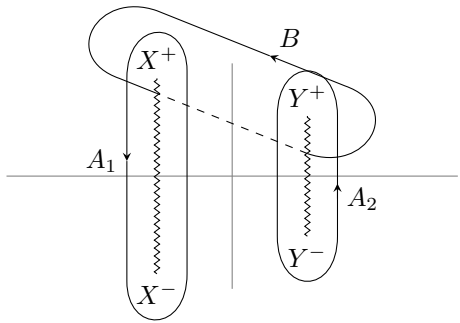}
  \caption{The periods $A_1$, $A_2$ and $B$.}
  \label{fig:ab-periods}
\end{figure}
Defining the periods $A_i$ and $B$ as
in figure~\ref{fig:ab-periods}, the differential $dp$ will be subject to
the following periodicity constraints
\begin{equation}
  \int_{A_i} dp = 0, \qquad \int_B dp = 2\pi n, \quad n \in \Integers.
\end{equation}
In the singular limit where $Y^+ \to X^+$, the $B$ period encircles
the endpoint $X^+$ once. To satisfy the periodicity condition, $dp$
has to have a simple pole of residue $-in$ at $X^+$. The expressions
for the global charges in terms of $dp$ now easily gives the giant
magnon dispersion relation~\cite{Okamura:2008jm}. Thus a single giant
magnon can be seen as a singular limit of an elliptic string state.

Hence we have two configurations corresponding to a giant magnon -- a
single condensate or two cuts sharing endpoints. As noted by Vicedo,
the two descriptions are related by $\grSL(2,\Integers)$
transformations, which exchange how the square root branch points are
connected to form cuts. Performing such a transformation on a general
finite gap solution gives a new solution, which corresponds to the
same solution of the equations of motion for the string, provided the
$A$- and $B$-periods of $dp$ are preserved. To ensure this we may need
to add extra condensates to the transformed solution.

\section{Finite size giant magnons}

In the previous section, we considered string solutions with one spin
infinitely large, corresponding, in conformal gauge, to an infinitely
long world sheet. For states with finite spin we expect the energy to
receive finite size corrections, analogous to the wrapping
interactions in gauge theory. At large coupling these corrections are
expected to be exponentially suppressed in the string length $J$
\cite{Ambjorn:2005wa}. This was confirmed for the magnon by Arutyunov
\textit{et~al.\@}~\cite{Arutyunov:2006gs} who derived an explicit
generalization of the giant magnon solution. They found corrections to
the dispersion relation of the form
\begin{equation}
  \delta\ve =  - \frac{16g}{e^2} \sin^3\frac{p}{2} e^{-2\frac{J}{4g\sin\frac{p}{2}}} + \dotsb.
  \label{eq:disp-rel-cor-AFZ}
\end{equation}
This result was later confirmed by Janik and
{\L}ukowski~\cite{Janik:2007wt}, using the thermodynamic Bethe ansatz
(TBA) and the related L\"uscher formulas for finite size corrections
in two-dimensional quantum field theories.

\section{Interacting magnons}

We can also imagine string states consisting of several
magnons. Hofman and Maldacena~\cite{Hofman:2006xt} showed that for
infinite $J$, magnons can scatter against each other with a
two-particle S-matrix which agrees with the $\grSU(2|2)$ S-matrix of
the gauge theory. The energy of these multi-magnon states is however
simply the sum of the individual magnon energies. In the gauge theory
we saw that for finite size states, interaction between the magnons
affected both the allowed sets of momenta and the total anomalous
dimension. Hence it would be interesting to see how the magnon
interactions effect the spectrum on the string side.

\subsection{Interacting finite size magnons as finite gap solutions}

The description of interacting string states is in general a very
complicated problem. From the point of view of the finite gap
equations, we noted previously that a single giant magnon is a two gap
solution. A state of $n$ giant magnons corresponds to a solution with
$2n$ cuts, and will hence be given as a function on a Riemann surface
of genus $2n-1$, which generally has to be expressed in terms of
hyperelliptic functions.

As noted above, the giant magnon in the infinite spin limit can be
regarded as a finite gap solution consisting of a single condensate,
resulting in a logarithmic resolvent, instead of the expected elliptic
form of a generic two gap solution. A state of several magnons is now
given by a number of such condensates. To see how this effects the
solution, we note that the only equation that induces interactions
between different cuts is eq.~\eqref{eq:P-discont-eq}, which describes
the discontinuity of $P(x)$ along a square root branch cut. But this
equation only involves square root cuts, and hence the many magnon
solution is simply a sum of the individual solutions, resulting in an
additive total energy, $E - J = \sum_i \ve_i$.

\begin{figure}
  \centering
  \includegraphics{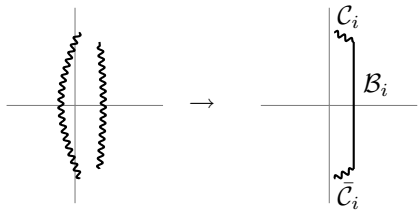}
  \caption{Finite gap configurations for a finite $J$ magnon as a two
    cut solution (left), and as a condensate with cuts at the ends
    (right). The two configurations are related by an
    $\grSL(2,\Integers)$ transformation.}
  \label{fig:recut}
\end{figure}

A giant magnon could also be described as a singular limit of a
two-cut solution. In this picture we expect finite size corrections to
arise when we let the endpoints of the two cuts be very close to each
other. Hence we will consider two square root branch cuts with a
separation of length $\delta$ between the endpoints. Performing an
$\grSL(2,\Integers)$ transformation on this configuration we end up
with a single condensate $\mathcal B_i$ with square root branch cuts
$\mathcal C_i$ and $\conj{\mathcal C}_i$ of length $\delta$ attached
at each end~\cite{Minahan:2008re}, as depicted in
figure~\ref{fig:recut}.

As an ansatz for the resolvent $G(x)$ of a set of finite size magnons
we write\footnote{In~\cite{Minahan:2008re} the density corresponding
  to this resolvent was derived using the integral equations
  from~\cite{Minahan:2006bd}. The same resolvent has also been used to
  calculate finite-size corrections to giant magnons in $\AdS_4 \times
  \mathrm{CP}^3$~\cite{Lukowski:2008eq}. A very similar construction
  was also used in~\cite{Gromov:2008} to calculate finite size
  corrections at the one-loop level.}.
\begin{align}
  G(x) &= \sum_i G_i(x), &
  G_i(x) &= -2i \log \frac{\sqrt{x-X^+_i} + \sqrt{x-Y^+_i}}{\sqrt{x-X^-_i} + \sqrt{x-Y^-_i}}.
\end{align}
The square root cuts of $G_i$ are such that the relative sign in the
numerator (denominator) changes when we cross $\mathcal C_i$
($\conj{\mathcal C_i}$).  As a simple check of this ansatz we can let
$Y^\pm_i \to X^\pm_i$ to recover the previous giant magnon resolvent.

Since we are interested in the leading order corrections, we make an
expansion by setting
\begin{equation}
  Y^\pm_i = X^\pm_i \pm i \delta_i e^{\pm i\phi_i} + \dotsb,
\end{equation}
where $\delta_i \ll 1$ is real. We also need to take into account the
back-reaction of $X^{\pm}_i$ by expanding it around $\delta_i$.
Calculating the asymptotic behavior as $x \to \infty$ and $x \to 0$
and solving the resulting equations iteratively to the second order in
$\delta_i$ by requiring that $p_i$ and $Q_i$ receive no corrections,
we get the correction\footnote{Since we want to consider only
  fundamental magnons, $Q_i / g$ will be put to zero. Dyonic magnons
  with $Q_i \sim g$ were treated in the same fashion in
 ~\cite{Minahan:2008re}. This case was also independently treated by
  Hatsuda and Suzuki~\cite{Hatsuda:2008gd}.}
\begin{equation}
  \delta\ve_i = - \frac{g\delta_i^2}{4} \cos\lrbrk{p - 2\phi_i}\sin\frac{p}{2} + \dotsb.
\end{equation}

So far there has been no contribution from the interactions between
the magnons. However we still have to determine the parameters
$\delta_i$ and $\phi_i$. We will do that by ensuring that the cuts
really correspond to square root branch cuts, requiring that they
satisfy \eqref{eq:P-discont-eq}. Thus we need to calculate $P(x +
i\epsilon) + P(x - i\epsilon)$ for $x$ on a cut $\mathcal C_i$ to the
leading order in $\delta_i$. Solving the resulting equation for
$\delta_i$, we get
\begin{equation}
  i\delta_i e^{i\phi_i} = 
  4 \lrbrk{X_i^+ - X_i^-} e^{-\frac{iE}{2g} \frac{X_i^+}{\lrbrk{X_i^+}^2 - 1} + i\pi m} \;
  \prod_{k \not = i} \frac{X_i^+ - X_k^-}{X_i^+ - X_k^+}
\end{equation}
Inserting this into the above expression for the correction to the
dispersion relation we get the final result
\begin{equation}
  \delta\ve_i =
  - \frac{16 g}{e^2} \sin^3 \frac{p_i}{2}\,  e^{-2 \frac{J}{\ve_i}} \;
    \prod_{k \neq i} \frac{\sin^2\frac{p_i + p_k}{4}}{\sin^2\frac{p_i - p_k}{4}}
    e^{-2 \frac{\ve_k}{\ve_i}}\,.
    \label{eq:multi-magnon-correction}
\end{equation}
The exponential suppression, as well as the pre-factor, agrees with
the one-magnon result in \eqref{eq:disp-rel-cor-AFZ}. The correction
is changed by the magnon interactions by a multiplicative factor of
order one, which is very similar in form to the two magnon scattering
phase in~\cite{Hofman:2006xt}.

\subsection{Interacting finite size magnons from sine-Gordon and TBA}

Interacting magnons can also be studied using the correspondence
between the equations of motions for an $\Reals \times \Sphere^2$
sigma model and those of the sine-Gordon model. Hofman and Maldacena
used this to calculate the magnon $S$-matrix. Klose and
McLoughlin~\cite{Klose:2008rx} considered periodic two-phase solutions
to the sine-Gordon equation. These solutions can be interpreted as
interacting finite two-magnon solutions. The resulting solutions
turned out to be elliptic, rather than hyperelliptic, which
considerably simplified the problem.  The corrections to the
dispersion relation perfectly agrees with the one calculated in the
previous sections.

The finite size corrections to the multi-magnon energy has also been
calculated using a generalization of the L\"uscher rules
\cite{Hatsuda:2008na}. Again the result agree with the finite size
calculation.

\section{Conclusions}

The spectrum of long operators in $\superN=4$ SYM and for large spin
states in the dual string theory is by now well known. The next step
in finding the full spectra is the understanding of the finite size
corrections to these states.

The giant magnon can be seen as a fundamental excitation on the string
world-sheet, dual to impurities propagating in the spin-chain
picture. At infinite $J$ the spectra of both theories consists of
states with magnons with arbitrary momenta (up to level-matching). The
magnons may scatter against each other, but the total energy is given
as a sum of the energies of the corresponding free magnons.

For finite $J$ the dispersion relation of the spin-chain impurities
receives wrapping corrections due to virtual excitations travelling
around the chain. In addition the allowed
momenta become quantized through a Bethe equation. The interaction
between the magnons gives essential contributions to both of these
corrections.

The dispersion relation of a single giant magnon gets exponential
corrections at finite $J$~\cite{Arutyunov:2006gs}. Again these
corrections stem from virtual excitations wrapping the world-sheet
\cite{Janik:2007wt}. The energy of a multi-magnon state is however not
the sum of the individual energies. In this paper, the finite-gap
calculation of the leading order contribution from magnon interactions
has been reviewed. The resulting order-one factors are related to the
two magnon S-matrix.

As in the gauge theory, we expect the momenta of finite $J$ giant
magnons to be quantized in terms of the magnon S-matrix. The exact
nature of this quantization remains an unsolved problem. The solution
would deepen the understanding of the relation between the
perturbative gauge and string theories.

\vspace{\baselineskip}\noindent
I would like to thank J.\@ Minahan and V.\@ Giangreco Marotta Puletti for
their valuable comments and discussions.

\end{document}